\documentclass{aastex631}

\shorttitle{an interior--orbital resonance in Jupiter--Io}
\shortauthors{Idini \& Stevenson}
\graphicspath{{./}{figures/}}

\begin{document}

\title{The gravitational imprint of an interior--orbital resonance in Jupiter--Io}

\correspondingauthor{Benjamin Idini}
\email{bidiniza@caltech.edu}

\author[0000-0002-2697-3893]{Benjamin Idini}
\affiliation{Division of Geological and Planetary Sciences, California Institute of Technology \\
1200 E California Blvd, MC 150-21 \\
Pasadena, CA 91125, USA}

\author[0000-0001-9432-7159]{David J. Stevenson}
\affiliation{Division of Geological and Planetary Sciences, California Institute of Technology \\
1200 E California Blvd, MC 150-21 \\
Pasadena, CA 91125, USA}



\begin{abstract}
At mid-mission perijove 17, NASA's Juno mission has revealed a $7\sigma$ discrepancy between Jupiter's observed high--degree tidal response and the theoretical equilibrium tidal response, namely the Love number $k_{42}$. Here, we propose an interpretation for this puzzling disagreement based on an interior--orbital resonance between internal gravity waves trapped in Jupiter's dilute core and the orbital motion of Io. We use simple Jupiter models to calculate a fractional correction $\Delta k_{42}$ to the equilibrium tidal response that comes from the dynamical tidal response of a $g$--mode trapped in Jupiter's dilute core. Our results suggest that an extended dilute core ($r\gtrsim0.7R_J$) produces an interior--orbital resonance with Io that modifies Jupiter's tidal response in $\Delta k_{42}\sim-11\%$, allowing us to fit Juno's $k_{42}$. In our proposed self--consistent scenario, Jupiter's dilute core evolves in resonant locking with Io's orbital migration, which allows the interior--orbital resonance to persist over geological timescales. This scenario requires a dilute core that becomes smoother or shrinks over time, together with a $_4^2g_1$ mode ($\ell,m,n=4,2,1$) with resonant tidal dissipation reaching $Q_4\sim1000$. Jupiter's dilute core evolution path and the dissipation mechanism for the resonant $_4^2g_1$ mode are uncertain and motivate future analysis. No other alternative exists so far to explain the $7\sigma$ discrepancy in Juno $k_{42}$. Our proposed interior--orbital resonance can be tested by Juno observations of $k_{42}$ tides raised on Jupiter by Europa as obtained at the end of the extended mission (mid 2025), and by future seismological observations of Jupiter's $_4^2g_1$ mode oscillation frequency.
\end{abstract}



\section{Introduction} \label{sec:intro}

In the traditional view of Jupiter's interior, an envelope of H-He fluid with small traces of heavier elements overlays a compact core of $10-20 M_E$ made entirely of elements heavier than H-He \citep{guillot2005interiors}. The traditional view follows from the simplest scenarios of planet formation via core accretion \citep{safranov1969evolution,perri1974hydrodynamic,mizuno1978instability,mizuno1980formation,pollack1996formation}, ignoring the disaggregation or dissolution of incoming planetesimals that recent work considers \citep{helled2017fuzziness,bodenheimer2018new}. An adiabatic Jupiter requires $10-20 M_E$ of heavy elements to attain its observed radius, irrespective of how those heavy elements distribute inside the planet. However, no direct geophysical evidence exists to justify the presence of a compact core over a smoother distribution of heavy elements. Opposing the traditional view, Juno recently obtained accurate zonal gravitational moments $J_{2\ell}$ that suggest that heavy elements distribute broadly along the radius rather than tightly concentrating near the center in a compact core \citep{wahl2017comparing, miguel2022jupiter} (Militzer et al., 2022).

In addition to $J_{2\ell}$, Juno also recently obtained accurate Love numbers that contain Jupiter's tidal response to the gravitational perturbation caused by the Galilean satellites. Intriguinly, the high--degre Love number $k_{42}$ observed by Juno is $7\sigma$ away from the hydrostatic $k_{42}$ calculated in a Jupiter model that fits the observed radius, $J_2$, and $J_4$ \citep{durante2020jupiter,wahl2020equilibrium}. The Love number $k_{42}$ represents the $\ell,m=4,2$ spherical harmonic term in Jupiter's tidal gravitational field normalized by the respective tidal forcing spherical harmonic term. The oblate figure of Jupiter introduces part of the tidal response to the $\ell=2$ tidal forcing into the $\ell=4$ tidal gravitational field, enhancing $k_{42}$ then compared to a hypothetical spherical Jupiter. In the case of tides raised on Jupiter by Io, only $7\%$ of $k_{42}$ corresponds to the tidal response to the $\ell=4$ tidal forcing and the remainding $93\%$ corresponds to the coupled $k_2$ ($\ell=m=2$) \citep{idini2022lost}.

Here, we use simple Jupiter models to propose an explanation to the $7\sigma$ disagreement between Juno and the hydrostatic $k_{42}$. We calculate a fractional correction to the hydrostatic $k_{42}$ introduced by the dynamical response of Jupiter's dilute core to tidal excitation. The hypothesized dilute core promotes static stability in the interior of Jupiter, allowing internal gravity waves to propagate and organize in normal modes of oscillation (i.e., $g$--modes) restored by buoyancy \citep{fuller2014saturn,mankovich2021diffuse}. Our proposal depends on an interior--orbital resonance between $g$--modes trapped in Jupiter's dilute core and the orbital motion of Io, a scenario that is highly unlikely to happen by pure chance. Accordingly, we invoke a state of resonant locking \citep{fuller2016resonance} to allow the required interior--orbital resonance to remain active over geological timescales.  

Our results support a previous suggestion that Jupiter's dilute core may extend as far as $\sim0.7R_J$ (Militzer et al., 2022), without allowing us to rule out a less extended dilute core. A future tight constraint to the extension of Jupiter's dilute core will lead to important consequences for our current understanding of the formation and evolution of gas giants. Extending outwards from a compact core, a narrow compositional gradient appears in some standard core accretion models \citep{helled2017fuzziness}. Double--diffusive convection could potentially broaden a narrow compositional gradient near the center by upward transport of heavy elements. On the other hand, convection over the age of Jupiter ($\sim 4.5$ Ga) promotes mixing in the envelope, potentially erasing compositional gradients that extend too far from the center \citep{muller2020challenge}. Perhaps an extended compositional gradient survives convective mixing only in the case of a "cold" formation process \citep{vazan2018jupiter}, which is not compatible with standard core accretion models \citep{muller2020challenge}. Alternatively, a head-on giant impact could disturb Jupiter after formation, leading to an extended compositional gradient resistant to convective mixing \citep{liu2019formation}. However, head–on giant impacts occur rarely and oblique giant impacts may not accomplish the desired heavy element distribution (Helled et al. 2021).

 \section{an interior--orbital resonance solves the Juno discrepancy\label{sec:dil_cor}}
 At mid--mission perijove 17, Juno registers a $7\sigma$ discrepancy in the observed $k_{42}$ when compared to the hydrostatic $k_{42}$ expected in a rotating Jupiter model that follows a density profile consistent with the observed radius and Juno zonal gravity up to $J_4$ \citep{durante2020jupiter,wahl2020equilibrium}. The hydrostatic Love number is  $k_{42}=1.743 \pm 0.002$ \citep{wahl2020equilibrium},  while the Juno observation is $k_{42}=1.289\pm0.063$ ($1\sigma$) \citep{durante2020jupiter}. The hydrostatic $k_{42}$ requires a fractional correction $\Delta k_{42}\approx-15\%$ to be reconciled at $3\sigma$ with the $k_{42}$ observed by Juno. 
 
 One part of the required fractional correction comes from the $-4\%$ effect introduced by $\ell=2$ $f$--mode dynamical tides on $k_2$ \citep{idini2021dynamical,Lai_2021,dewberry2022dynamical} and coupled by the oblate figure of the planet into the $\ell=4$ gravitational field \citep{idini2022lost}. The $\ell=4$ $f$--mode dynamical tide produces a negligible effect on $k_{42}$, thus the remaining $\Delta k_{42}\approx-11\%$ comes from additional dynamical effects related to the tidal response of nonfundamental modes to the $\ell=4$ tidal forcing. 

In the rest of this section, we present a self--consistent scenario where resonantly enhanced internal gravity waves trapped in Jupiter's extended dilute core produce the tidal gravity required to reconcile the $7\sigma$ Juno discrepancy.

 Instead of a compact traditional core, gas giant planets most likely host a dilute core where the abundance of heavy elements changes with depth \citep{fuller2014saturn}. Recent observations of normal oscillations in Saturn's rings confirm this picture in Saturn \citep{mankovich2021diffuse}, whereas high--degree zonal gravitational coefficients observed by Juno suggest an analogous situation in Jupiter \citep{wahl2017comparing}. Contrary to a homogeneous envelope, a dilute core permits the propagation of internal gravity waves (i.e., waves restored by buoyancy) that organize in normal modes of oscillation called $g$--modes. The oscillation frequency of $g$--modes spans a wide frequency range that includes the forcing tidal frequency of the Galilean satellites. Consequently, in a gas giant planet hosting a dilute core, the orbital motion of the satellites may resonate with internal gravity waves trapped in the dilute core, leading to what we denominate an interior--orbital resonance. As we show in Section \ref{sec:dynk42}, internal gravity waves trapped in the dilute core fail at producing enough gravity to explain Juno's observation when out of resonance.
 
 The tidal excitation of $g$--modes trapped inside Jupiter's dilute core produces a fractional correction $\Delta k_{42}$ to the hydrostatic Love number defined as
\begin{equation}
    \Delta k_{42} =  \frac{k_{42}}{k_{42}^{\textnormal{(hs)}}}-1 \textnormal{,}
    \label{eq2:corrections}
\end{equation}
where $k_{42}$ is the dynamical Love number (Appendix \ref{app:dynamic}) and  $k_{42}^{\textnormal{(hs)}}\approx0.12$ the hydrostatic Love number \citep{idini2022lost}, both calculated in an $n=1$ polytrope.The $n=1$ polytrope closely approximates the equation of state of H-He, the elements that dominate the composition of gas giant planets \citep{stevenson2020jupiter}. We use perturbation theory to obtain the $\Delta k_{42}$ required by Juno rather than directly trying to fit the observed $k_{42}$. A polytropic equation of state and our perturbative approach greatly simplify an otherwise much more complicated numerical procedure without compromising the generality of our results.

A gyrotidal effect on Jupiter couples Love numbers to rotation \citep{idini2022lost,dewberry2022dynamical} and introduces an additional complication when comparing Equation (\ref{eq2:corrections}) to the fractional correction required by Juno. Due to the gyrotidal effect, an interior--orbital resonance only affects the contribution to $k_{42}$ that comes from the $\ell=4$ tidal forcing, which represents only a small fraction ($7\%$) of the total $k_{42}$ associated to Io \citep{idini2022lost}; the contribution to $k_{42}$ from the nonresonant coupled response to the $\ell=2$ forcing remains hydrostatic. Consequently, we calculate the $g$--mode fractional dynamical correction $\Delta k_{42}$ as
 \begin{equation}
     \Delta k_{42}\simeq 0.07\left(\frac{4\pi}{9k_{42}^{\textnormal{(hs)}}}\right)\left(\frac{\mathcal{Q}^2}{\omega_{g}^2-(2C\Omega+\omega)^2}\right)\textnormal{,}
     \label{eq:dlai}
 \end{equation}
 where the numerical factor 0.07 accounts for Jupiter's gyrotidal effect, $\omega_g$ is the $g$--mode frequency, $\omega$ is the tidal frequency, $\mathcal{Q}$ is the dimensionless coupling integral that represents how well the tidal forcing couples to the eigenvector of the normal mode, and $C$ is the dimensionless amplitude of the first--order correction to the mode frequency due to Jupiter's rotation rate $\Omega$,
 \begin{equation}
     \mathcal{Q} = \frac{4\pi}{MR^\ell}\int_0^R r^{\ell+2}\delta\rho^*dr\textnormal{,}
     \label{eq:integral_coupling}
 \end{equation}
 \begin{equation}
     C = \frac{4\pi}{MR^2}\int_0^R\left(2\xi_r\xi_\perp + \xi_\perp^2\right)\rho r^2dr\textnormal{,}
     \label{eq:rotational_coeff}
 \end{equation}
 where $\delta\rho^*$ is the complex conjugate of the Eulerian densitiy perturbation of the mode, $\xi_r$ is the radial eigenfunction of the mode, and $\xi_\perp$ is the horizontal eigenfunction of the mode. At $\ell=4$, rotation produces only a small shift on the $g$--mode frequency (e.g., $C\leq0.05$ for models shown in this paper), thus resonances still appear roughly at $\omega_g\approx\omega$. At a given satellite, the tidal frequency at $m=2$ follows
 \begin{equation}
     \omega=-2(\Omega-\omega_s)\textnormal{,}
 \end{equation}
where $\omega_s$ is the orbital frequency of the satellite that depends on the semimajor axis according to $\omega_s\propto a^{-3/2}$.

To fit the Juno observation, we require to approach an interior--orbital resonance from the resonance branch that produces a negative fractional correction to $k_{42}$. According to Equation (\ref{eq:dlai}), we obtain the required negative fractional correction only when $|\omega|>\omega_g$. 

Pure chance hardly favors an interior--orbital resonance in the hypothetical scenario that the Galilean satellites are randomly placed in orbit around Jupiter. Tidal torques constantly force the satellite to migrate outwards, increasing the semimajor axis $a$ and $\omega$ over time. Likewise, the dilute core structure evolves over time due to convection (either overturning convection or double--diffusive convection), correspondingly changing the $g-$mode frequency. Resonant locking (Appendix~\ref{app:resonant}) allows us to circumvent the invocation of a historical coincidence to explain why these two frequencies match in present days. Resonant locking is an equilibrium state where the evolution of the dilute core roughly matches the orbital migration of the forcing satellite \citep{fuller2016resonance}. 

\begin{figure}[ht!]
    \centering
    \plotone{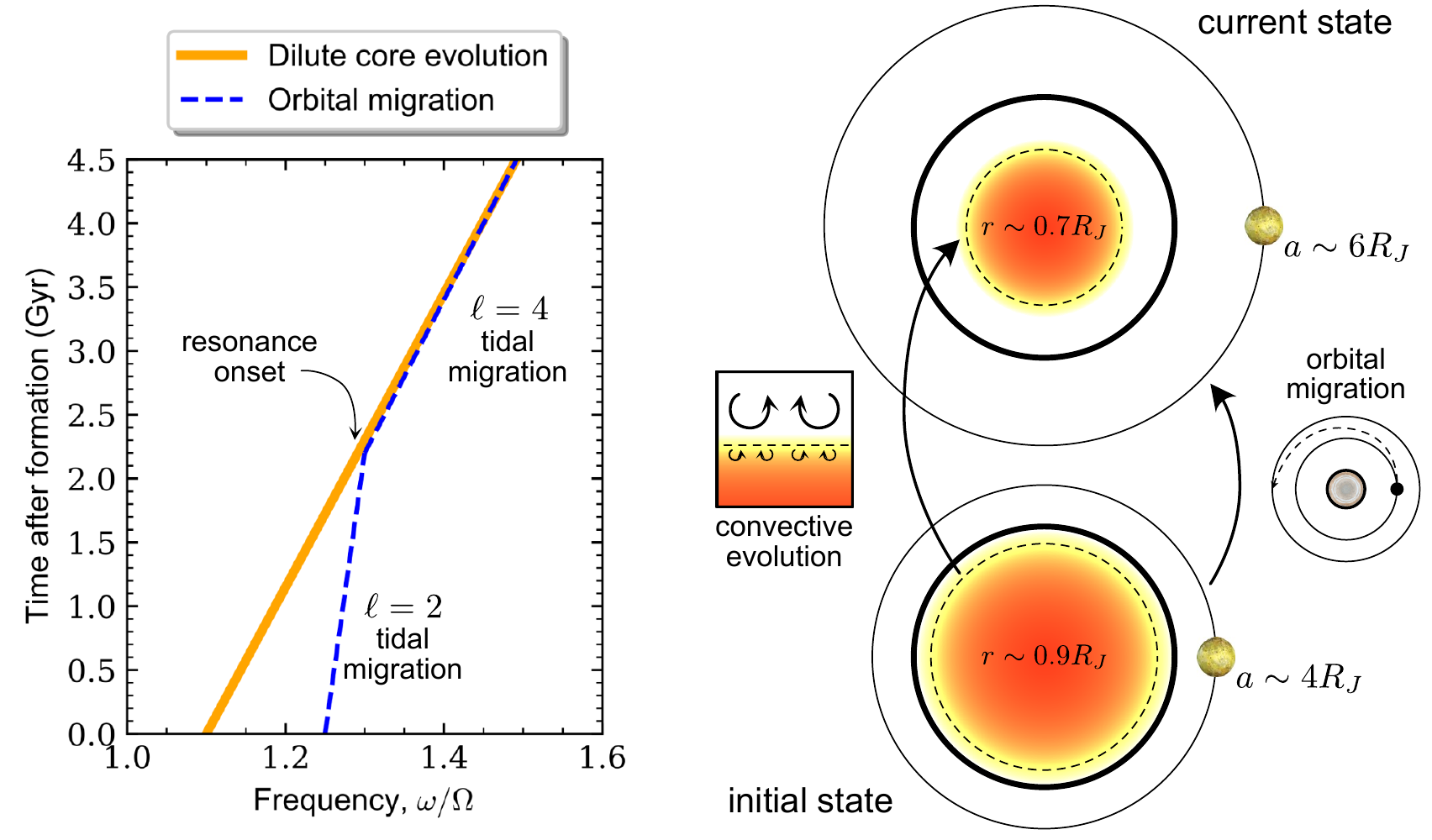}
    \caption{Pictographic orbital migration and dilute core evolution required to explain Juno's $k_{42}$ using a core--orbital resonance. The dilute core forms with a $g$--mode frequency lower than the satellite tidal frequency. Initially out of resonance, the satellite migrates outward at an almost negligible rate. After some uncertain time, the dilute core evolves into a locked interior--orbital resonance, increasing the rate of satellite orbital migration. \label{fig:evolution}}
\end{figure}

We propose the following scenario to maintain an interior--orbital resonance over geological timescales that produces a negative fractional correction to $k_{42}$ and fits Juno's observation (Fig. \ref{fig:evolution}). An initially nonresonant extended dilute core (i.e., $r\gtrsim0.7R_J$) hosts a $g$--mode that evolves into higher frequency due to some kind of convection. In this initial stage, the Galilean satellites slowly migrate outward, expanding their orbits due to dissipation associated with $\ell=2$ tides. Eventually, the $g$--mode frequency encounters a resonance with Io, the Galilean satellite of lowest tidal frequency. The onset of the resonance increases tidal dissipation at $\ell=4$, accelerating Io's orbital migration until achieving a state of resonant locking (Section~\ref{sec:migration}). The system could remain in resonant locking for geological timescales until present day, assuming that the $g$--mode evolved faster toward higher frequency than the orbital migration of Io while they were out of resonance.

In the following, we evaluate the plausibility of the scenario shown in Fig. \ref{fig:evolution}. In Section \ref{sec:model}, we describe our simple models of Jupiter's dilute core. In Section \ref{sec:dynk42}, we show that an interior--orbital resonance in Jupiter--Io prefers an extended dilute core. In Section \ref{sec:migration}, we calculate the tidal dissipation required to attain $\ell=4$ resonant locking. 

  \section{Jupiter dilute core models \label{sec:model}}
 
\begin{figure}[t]
    \centering
    \plotone{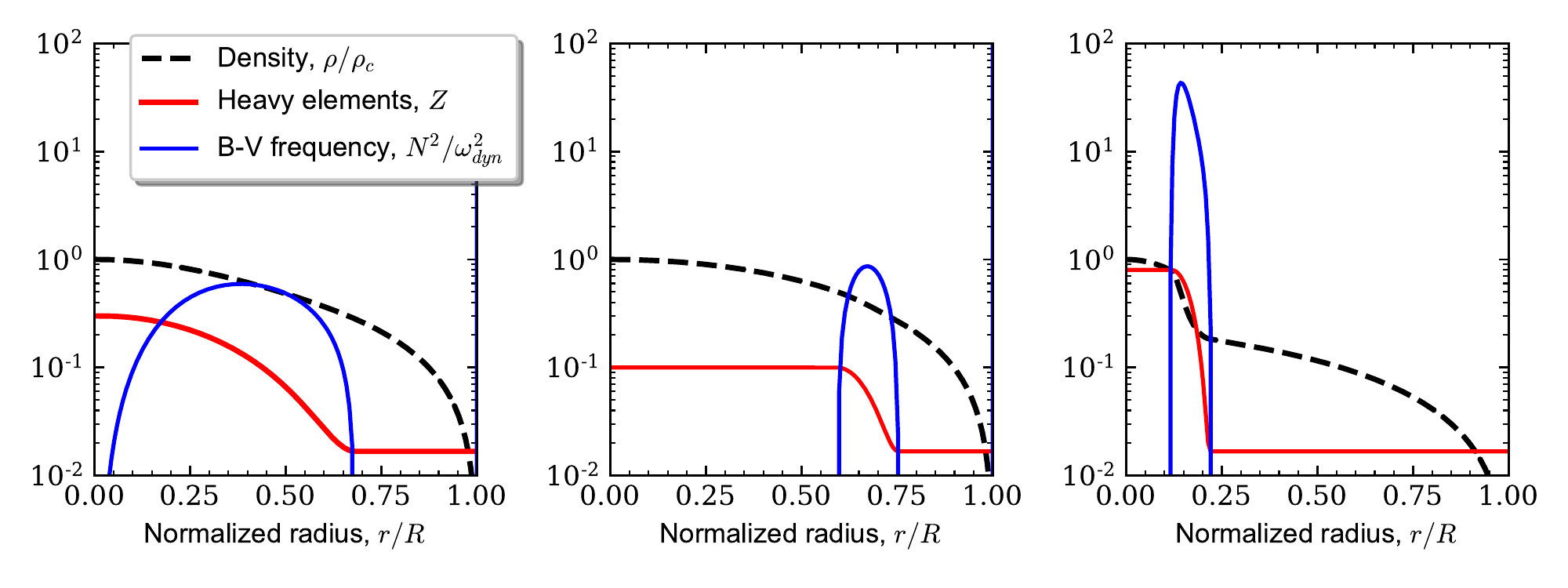}
    \caption{Jupiter interior models with a dilute core defined as a gradient in the enrichment of heavy elements. (a) A wide dilute core with a smooth compositional gradient similar to that proposed in \cite{debras2019new}, (b) a narrow dilute core with a sharp compositional gradient similar to that proposed in Militzer et al. (2022), and (c) a compact dilute core with a sharp compositional gradient constrained to a central region, similar to a traditional core. The density profile is normalized by a central density $\rho_c$ equal to $5.8$, $4.6$, and $23.1$ g cm$^{-3}$, respectively. The enrichment of heavy elements $Z$ corresponds to the mass fraction of elements heavier than H and He. The Brunt-Vaisala (B-V) frequency is normalized by Jupiter's dynamical frequency $\omega_{dyn}^2= \mathcal{G}M_J/R_J^3$.\label{fig:interior}}
\end{figure}

The density $\rho $ of a mixture of H-He fluid and heavy elements as a function of pressure $p$ (Fig.~\ref{fig:interior}) can be obtained from volume additivity of the individual constituents (Appendix \ref{app:mixture})
\begin{equation}
    \frac{1}{\rho} = (1-Z)\left(\frac{K}{p}\right)^{1/2} + \frac{Z}{\rho_z}\textnormal{,}
    \label{eq:specificvol}
\end{equation}
where $\rho_z$ is the density of heavy elements, $Z$ the fraction of mass corresponding to elements heavier than H-He, and $K\approx2.1\cdot10^{12}$ cgs represents the bulk elastic properties of the H-He fluid for a cosmic abundance of He. For simplicity, we again adopt an $n=1$ polytrope to approximately represent the response of the H-He fluid. Provided $Z$ is not large, the solution to Equation (\ref{eq:specificvol}) is not much different from the $n=1$ polytrope. For example, the dilute core models of Militzer et al. (2022) closely follow an $n=1$ polytrope with a slightly different effective $K$ \citep{stevenson2020jupiter}.

For convenience, we parameterize the enrichment of heavy elements in a dilute core of width $L_c$ and inner radius $x_{ic}$ following \citep{fuller2014saturn}
\begin{equation}
    Z(x) = Z_e + (Z_c - Z_e)\sin^2\left(\frac{\pi(x_{ic}+L_c-x)}{2L_c}\right)
    \label{eq:Zx}
\end{equation}
where $Z_e$ and $Z_c$ represent the enrichment of heavy elements in the envelope ($x>x_{ic}+L_c$) and at center of the planet ($x<x_{ic}$), respectively. The normalized radius follows $x=kr$, where $k^2 = 2\pi\mathcal{G}/K$. The enrichment of heavy elements decays from $Z_c$ to $Z_e$ following the $\sin^2$ function along the dilute core width (Fig.~\ref{fig:interior}).

The total mass of heavy elements in our models range from 18 to 25 $M_E$, in agreement with estimates provided by other interior models \citep{guillot2005interiors}. We construct our dilute core models by fixing the enrichment of heavy elements in the envelope to $Z_e=0.0167$, the value observed by two independent instruments in the Galileo entry probe. The free parameters in our models are $Z_c$ and $L_c$. The presence of heavy elements shrinks Jupiter's radius when compared to a planet made of pure H-He fluid. We set the parameter $x_{ic}$ to fit the target planetary radius $R = 3/k$ in all of our models. The target planetary radius ultimately constrains the total mass of heavy elements.

A compositional gradient in heavy elements introduces static stability to the interior of Jupiter, represented by the Brunt-Vaisala frequency (Fig.~\ref{fig:interior}),
\begin{equation}
    N^2 = g\left(\frac{1}{\gamma}\frac{\partial\ln p}{\partial r} - \frac{\partial\ln \rho}{\partial r}\right)\textnormal{.}
    \label{eq2:N}
\end{equation}
In calculating $N^2$, we set the first adiabatic index to $\gamma=2$, which represents the adiabatic response of an $n=1$ polytrope. Thermal effects could possibly modify the static stability, but those effects are small for reasonable central temperatures \citep{mankovich2021diffuse} and neglected here.

 \begin{figure}[t]
 \centering
    \includegraphics[width=0.5\textwidth]{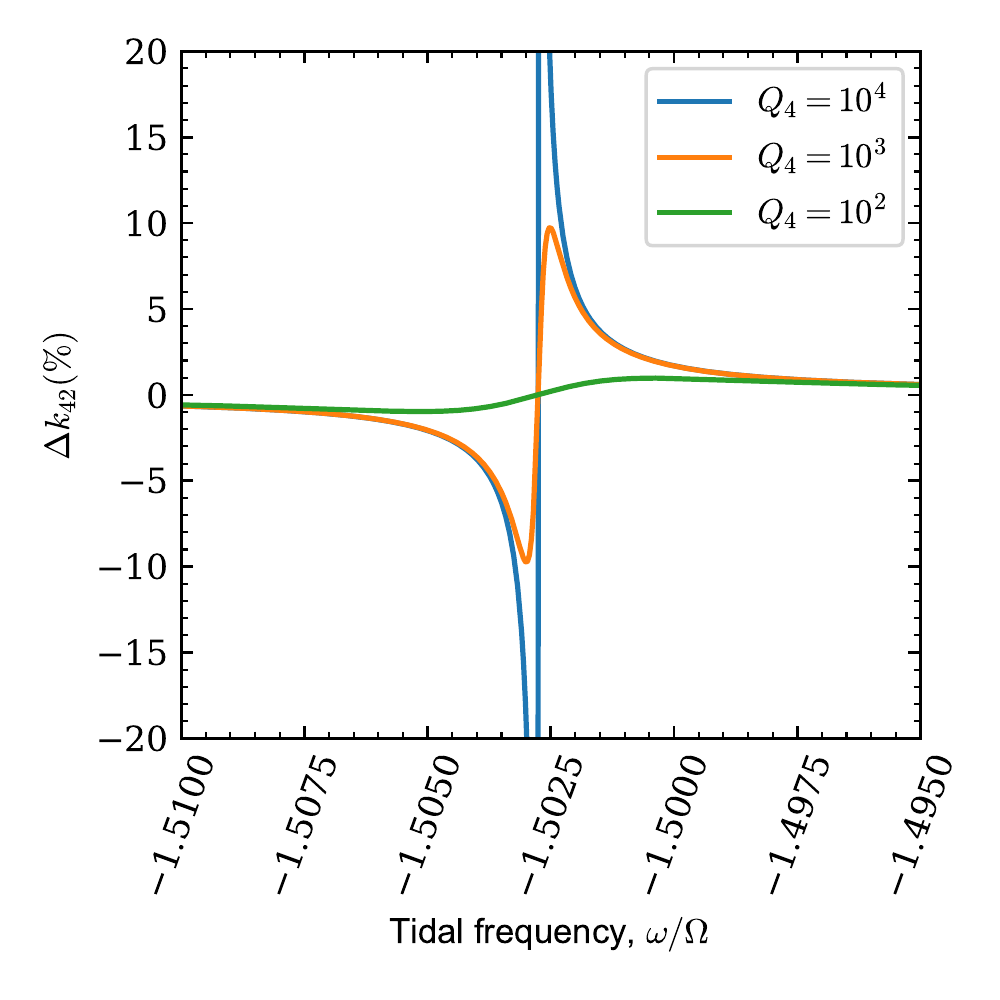}
    \caption{Fractional correction to the hydrostatic Love number $k_{42}$ as a function of tidal frequency and dissipation $Q_4$. The dilute core model (Fig.~\ref{fig:interior}b) produces a $g_1$--mode in resonance with the tidal frequency $\omega\approx -1.5\Omega$, a forcing frequency close to Io's tidal frequency $\omega_{Io}\approx-1.53\Omega$.\label{fig:resonance}}
\end{figure}

 \section{Tidal excitation of the dilute core \label{sec:dynk42}}
 In this section, we calculate the  fractional dynamical correction $\Delta k_{42}$ produced by the tidal excitation of our simple models of Jupiter's dilute core.  We restrict our analysis to the first--order $\ell,m=4,2$ $g$--mode (i.e., $_4^2g_1$), which produces a better coupling with the respective tidal forcing compared to higher order $g$--modes (i.e, $_4^2g_n$ modes, where $n>1$).
  We obtain $\omega_g$, $\mathcal{Q}$, and $C$ in Equation (\ref{eq:dlai}) using the stellar oscillations code GYRE \citep{townsend2013gyre} applied to our simple Jupiter models. We observe a fractional dynamical corection capable of reaching the required $\Delta k_{42}\approx-11\%$ only when the $_4^2g_1$ mode frequency approaches a resonance with the tidal frequency of the Galilean satellites (i.e., Fig.~\ref{fig:resonance}). The resonant model in Fig.~\ref{fig:resonance} produces a $\sim 1$ m radial displacement of Jupiter's outer boundary in order to obtain the required gravitational signal. 
  
  The amount of tidal dissipation in the $_4^2g_1$ mode is limited to $Q_4\gtrsim 1000$, otherwise internal gravity waves are damped below the amplitude required to explain Juno $k_{42}$ (Fig.~\ref{fig:resonance}). Our account of dissipation represents only a rough estimate, following a substitution in Equation (\ref{eq:dlai}) of the $g$--mode frequency,
 \begin{equation}
    \tilde{\omega}_g = \omega_g\left(1+\frac{i}{2\pi Q}\right)\textnormal{.}
    \label{eq:linearQ}
 \end{equation}
Our simplified dissipation model is equivalent to introducing the term $\mathbf{v}/(2\pi Q)$ in the right-hand side of the equation of motion (Equation (\ref{eq1:momentum})), a term that accounts for a convenient mathematical representation of frictional damping \citep{ogilvie2009tidal}.

\begin{figure}[t]
 \centering
    \includegraphics[width=0.5\textwidth]{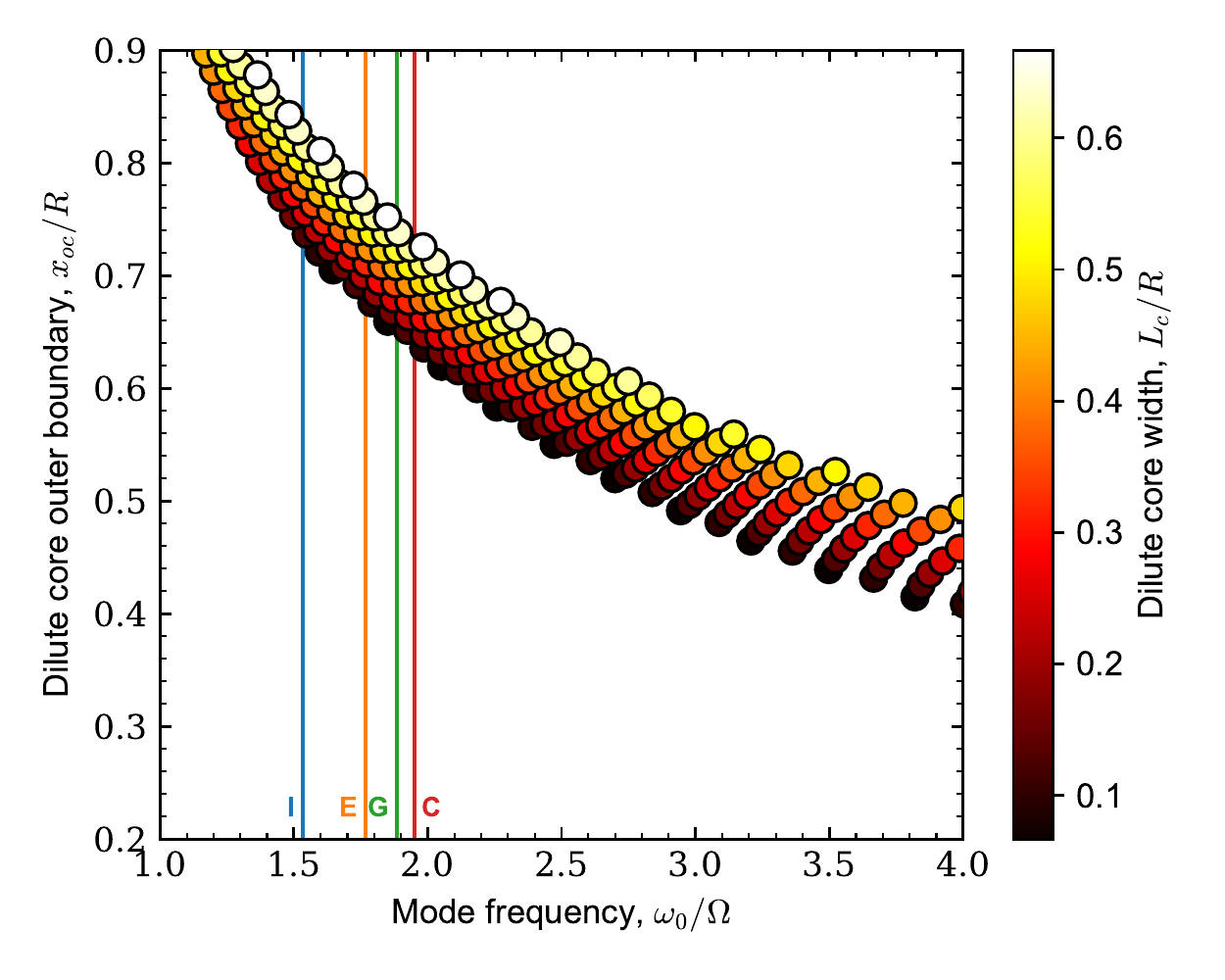}
    \caption{ The $_4^2g_1$ mode frequency of several dilute core models. The vertical lines represent the tidal frequency $|\omega|$ of Io, Europa, Ganymede, and Callisto, from left to right, respectively. \label{fig:models}}
\end{figure}

 Our results indicate that an extended dilute core (i.e., extending as far as $\gtrsim0.7R_J$) produces a $_4^2g_1$ mode frequency that resonates with the tidal frequency of the Galilean satellites (Fig.~\ref{fig:models}). Our compact dilute core models produce a $_4^2g_1$ mode frequency considerably above the tidal frequency of the Galilean satellites $\omega<2\Omega$, thus far from resonance. We cannot rule out the possibility of a resonance in a compact core with higher order $_4^2g_n$ modes (i.e., $_4^2g_n$ modes with extra radial nodes, where $n>1$) of worse tidal coupling because the $_4^2g_n$ mode frequency diminishes with increasing order $n$ \citep{aerts2010asteroseismology}. Consequently, we cannot constrain the extension of the dilute core purely based on the identification of an interior--orbital resonance at certain frequency. However, the lower tidal coupling of higher order $_4^2g_n$ modes leads to narrower resonances with lower saturation points, making the establishment of resonant locking harder when compared to the $_4^2g_1$ mode (Appendix~\ref{app:resonant}).

Can the dilute core trap a satellite in an interior--orbital resonance at $\ell=2$? In general, the $g$--mode frequency scales down with lower degree $\ell$ \citep{aerts2010asteroseismology}. At low degree $\ell$ and large number of radial nodes $n$, we can write
\begin{equation}
    \omega_{\ell,n} \simeq \frac{\sqrt{\ell(\ell+1)}}{\pi n}\int_0^{R} \frac{N}{r}dr \textnormal{.}
    \label{eq:gmode}
\end{equation}
According to Equation (\ref{eq:gmode}), the first--order $\ell=2$ $g$--mode (i.e., $_2^2g_1$) approximately oscillates with a  frequency that is roughly $1/\sqrt{3}$ times lower than the $_4^2g_1$ mode frequency. The $g$--mode spacing determined this way represents only a rough estimate because the $_2^2g_1$ and $_4^2g_1$ modes are far from the high--$n$ asymptotic limit. Considering a current $\ell=4$ interior--orbital resonance at $\omega_g\approx1.5\Omega$, the corresponding $_2^2g_1$ mode frequency is $\omega_g\approx0.87\Omega$, which is far from resonant. The $\sqrt{3}$ factor provides enough spacing between the mode frequency at different $\ell$ for the dilute core to evolve into a $\ell=4$ resonance without interfering with the $_2^2g_1$ mode. 

\begin{figure}[t]
 \centering
    \includegraphics[width=0.5\textwidth]{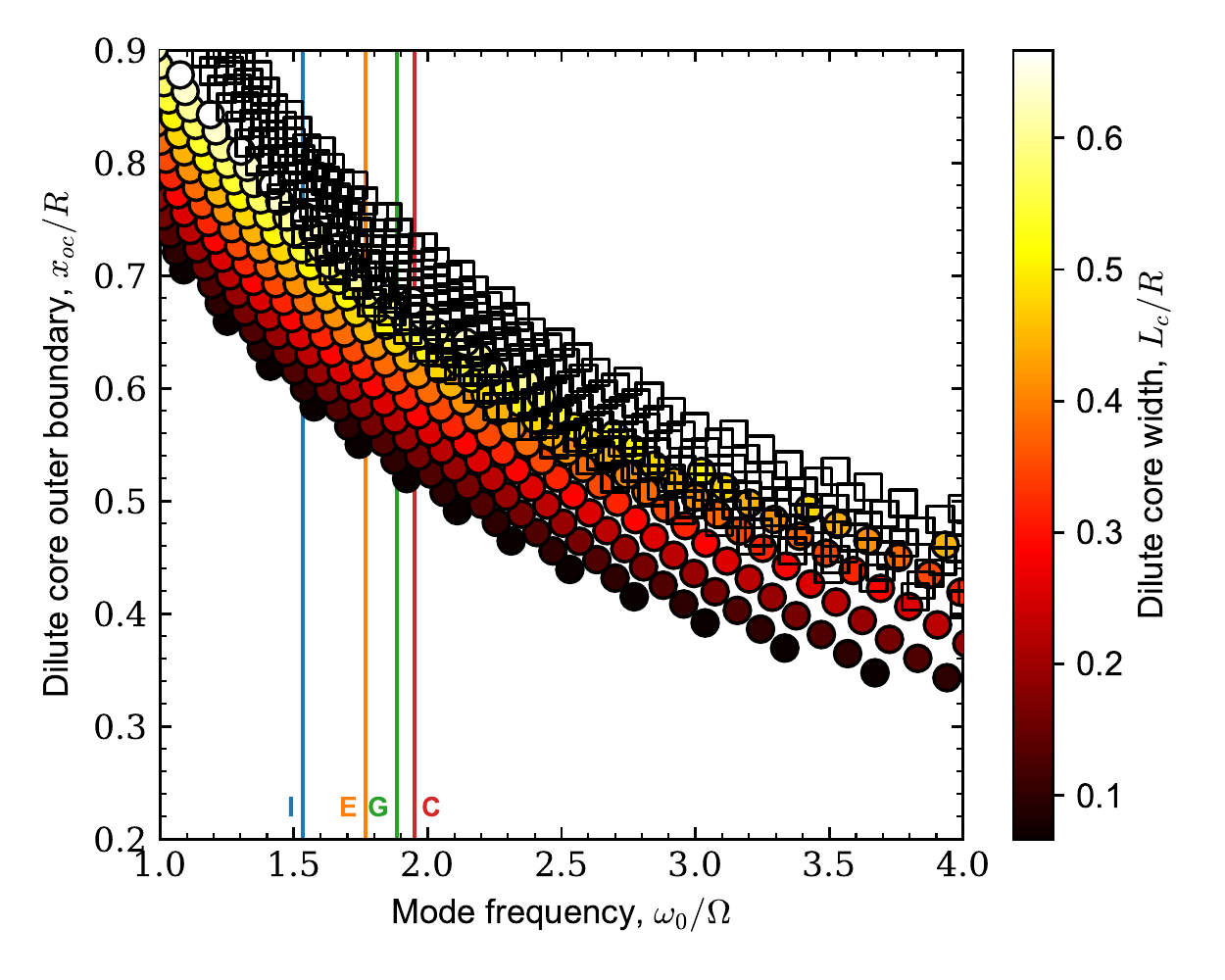}
    \caption{Same as Fig.~\ref{fig:models} but for the $_2^2g_1$ mode frequency. The empty rectangles show the higher--frequency $_4^2g_1$ mode of the same dilute core models. \label{fig:modelsk2}}
\end{figure}

Our dilute core models could potentially produce an $\ell=2$ interior--orbital resonance for realistic dilute core geometries (Fig.~\ref{fig:modelsk2}), but Juno $k_2$ observation argues against such scenario in present times. Juno observed $k_2=0.565\pm0.006$ at perijove 17 \citep{durante2020jupiter}, which is in close agreement with a rotating gas giant out of any significant interior resonance at $\ell=2$ \citep{idini2021dynamical,Lai_2021,dewberry2022dynamical}. The Juno $k_2$ could only admit a small near--resonance effect of a few percent, which would not be enough to explain the $\Delta k_{42}\approx-11\%$ required to reconcile the $7\sigma$  discrepancy in $k_{42}$. 

\section{Constraints to tidal dissipation imposed by resonant locking\label{sec:migration}}

In this section, we show that the $\ell=4$ tidal bulge can dominate the orbital migration of a satellite over the migration produced by $\ell=2$ tides, a requirement to establish a long--lived $\ell=4$ interior--orbital resonance via resonant locking.

The orbit of a satellite evolves in time due to a gravitational torque $\Gamma$ exerted on the satellite by the tidal bulge raised on the planet. In particular, the semimajor axis of a satellite of mass $m_s$ evolves as $\dot{a} \propto\Gamma$, with the gravitational torque following \citep{murray1999solar}
\begin{equation}
    \Gamma_\ell = -m_s \frac{\partial (_\ell^m\Theta^{'})}{\partial \alpha}\textnormal{,}
\end{equation}
where $\alpha$ is the angle that the tidal bulge lags  behind the gravitational pull of the satellite due to tidal dissipation. The tidal bulge produces an external potential $\Theta$ proportional to $r^{-(\ell+1)}\mathcal{P}_\ell(\cos\alpha)$, which is a solution to Laplace's equation. The $\ell=2$ external potential evaluated at the position of the satellite can be written as 
\begin{equation}
    _2^2\Theta^{'} = \frac{\mathcal{G}m_s}{a} \left(\frac{s}{a}\right)^5 k_2 \mathcal{P}_2(\cos\alpha_2)\textnormal{,}
    \label{eq:ext2}
\end{equation}
where $s$ is the average planetary radius.

The oblate figure of a rotating planet introduces significant changes to the structure of the tidal gravitational field. The $\ell=4$ tidal gravitational field acquires a new coupled nonresonant term from the $\ell=2$ tidal forcing, in addition to the term related to the direct response to the $\ell=4$ tidal forcing \citep{idini2022lost}. The external potential induced by the coupled nonresonant term and the direct term evaluated at the satellite position are, respectively,
\begin{equation}
    _4^2\Theta^{'} \sim \frac{5q}{\pi^2}\frac{\mathcal{G}m_s}{a} \left(\frac{s}{a}\right)^7 k_2 \mathcal{P}_4(\cos\alpha_4)\textnormal{,}
    \label{eq:ext2A}
\end{equation}
\begin{equation}
    _4^2\Theta^{'} = \frac{\mathcal{G}m_s}{a} \left(\frac{s}{a}\right)^9 k_{42} \mathcal{P}_4(\cos\alpha_4)\textnormal{,}
    \label{eq:ext2B}
\end{equation}
where $q$ is the dimensionless rotational parameter of a planet of mass $M$,
\begin{equation}
    q = \frac{\Omega^2s^3}{\mathcal{G}M}\textnormal{.}
\end{equation}
We explicitly indicate different lag angles $\alpha_\ell$ because dissipation can vary depending on the proximity between the tidal frequency and the mode frequency. The angle 
$\alpha$ increases when the mode frequency approaches a resonance with the tidal frequency. In Equation (\ref{eq:ext2A}), the relevant mode frequency is the $\ell=2$ $f-$mode, which is much higher than the tidal frequency of the Galilean satellites and hence leads to a comparatively small $\alpha_4$. In Equation (\ref{eq:ext2B}), on the other hand, the relevant mode frequency is the $_4^2g_1$ mode, which can resonate with Io's tidal frequency, leading to a comparatively large $\alpha_4$.

Assuming that $\alpha$ remains a small angle in all cases, we compare the evolution of the satellite's semimajor axis promoted by the $\ell=2$ tidal torque and the two $\ell=4$ tidal torques. The coupled nonresonant term and the direct term produce significantly different migration rates, respectively,
\begin{equation}
    \frac{\dot{a}_2}{\dot{a}_4}\sim\frac{3\pi^2}{50q}\left(\frac{Q_4}{Q_2}\right)\left(\frac{a}{s}\right)^{2}\textnormal{,}
\end{equation}
\begin{equation}
    \frac{\dot{a}_2}{\dot{a}_4}=\frac{3}{10}\left(\frac{k_2}{k_{42}}\right)\left(\frac{Q_4}{Q_2}\right)\left(\frac{a}{s}\right)^{4}\textnormal{,}
    \label{eq:q4}
\end{equation}
where we use $2\alpha_\ell\approx Q^{-1}_\ell$, and $Q_\ell$ is the tidal dissipation at a given degree $\ell$. 

We require $Q_4\lesssim1000$ for the $\ell=4$ external potential in Equations (\ref{eq:ext2A}) and (\ref{eq:ext2B}) to  significantly contribute to Io's orbital migration (i.e., $\dot{a}_4\gtrsim \dot{a}_2$), under the reasonable parameters $q\sim0.1$, $a/s\approx6$, $k_2/k_{42}\approx0.4$, and $Q_2\sim 10^5$. The nonresonant potential in Equation (\ref{eq:ext2A}) hardly produces the low $Q_4$ required in the analysis above, and therefore produces a negligible contribution to orbital migration. However, the external potential in Equation (\ref{eq:ext2B}) can reach the relatively low $Q_4$ required as a result of a resonance between Jupiter's $_4^2g_1$ mode and the tidal frequency associated to Io \citep{fuller2016resonance}. In this case, the dissipation in the required low $Q_4$ describes kinetic energy leaving the $_4^2g_1$ mode mostly due to energy cascading into higher degree $_\ell^2g_1$ modes ($\ell>4$) sustained by mode coupling. Particularly for internal gravity waves, dissipation mostly occurs in the turbulent breaking of oscillations at short wavelengths. Currently, no widely accepted explanation exists for the origin of the tidal $Q$ inside gas giant planets, which should be the subject of future work. 
 
The $Q_4$ required to guarantee dominance of the $\ell=4$ tidal torque (upper bound) is of the same order of magnitude than the $Q_4$ required to obtain the necessary gravitational effect to explain Juno's $k_{42}$ (lower bound), namely $Q_4\sim 1000$. The coincidence suggests a limitation in our proposal, which can be resolved by the following possibilities. Firstly, the upper bound extends upward if $Q_2$ is higher than what has been assumed so far, a possibility  we discuss in further detail in Section \ref{sec:Qjup}. Secondly, the linear approach implied in Fig.~\ref{fig:resonance} and Equation (\ref{eq:linearQ}) may not adequately represent tidal dissipation to the extend required in this problem. Finally, we should consider the fortuitous possibility that the current dissipation lies at $Q_4\sim1000$. Notwithstanding these concerns, there is no known alternative to explain the $k_{42}$ observed by Juno.

\section{Discussion}

\subsection{Previous estimates of Jupiter's tidal dissipation \label{sec:Qjup}}

A century of astrometric observations of the Galilean satellites ephemerides constrains Jupiter's recent dissipation to $k_2/Q_2\approx 10^{-5}$, which leads to $Q_2\approx6\cdot10^{4}$ for a realistic $k_2\approx 0.6$ \citep{lainey2009strong}. The analysis assumes the same Jupiter $k_2/Q_2$ at each satellite (i.e., no dynamical tides) and neglects any contribution to orbital migration from $k_{42}/Q_4$ or other higher degree tides. A different set of assumptions may lead to a much larger Jupiter's $Q_2$ than previously considered, which would allow us to reconcile our upper ($Q_4\gtrsim1000$) and lower bound ($Q_4\lesssim1000$) on $Q_4$. In particular, our scenario suggest a resonant $k_{42}/Q_4 \sim 10^{-3}$ that in principle could contribute to set Io's ephemerides as registered in the last century, allowing $k_2/Q_2$ to assume a lower value while maintaining the same secular migration rate $\dot{a}\sim10$ cm/yr. A small $k_2/Q_2$ is in agreement with the nonresonant $Q_2\sim10^6$ predicted by dynamical tide theories of gas giant planet dissipation \citep{ogilvie2004tidal}. Further analysis is required to test this hypothesis.

In our proposed scenario, an interior--orbital resonance vigorously pushes Io outward against the mean-motion resonance with the other Galilean satellites. A short-lived eccentricity cycle \citep{ojakangas1986episodic} can explain the astrometic suggestion that Io has migrated inward in the last century without compromising our proposal of outward migration at a longer secular timescale. The width of the resonance required to maintain resonant locking is $\delta\omega\sim 0.3$ $\mu$Hz (Fig.~\ref{fig:resonance} with Jupiter's spin--rate $\Omega\approx 170$ $\mu$Hz). Astrometry suggest that Io's orbital migration is $\dot{\omega}_s/\omega_s\sim 10^{-11}$/yr in the last century \citep{lainey2009strong}. When we apply the currently observed migration rate over a period of 100 years, Io's migration only changes the tidal frequency by $\Delta\omega_s\sim 4\cdot10^{-8}$ $\mu$Hz (Io's orbital frequency $\omega_s\approx 41$ $\mu$Hz), a quantity much smaller than the width of the resonance. Io  would require more than $\sim 1$ Gyr to migrate out of the proposed resonance at the migration rate observed by \cite{lainey2009strong}.

\subsection{Evolution of Jupiter's dilute core\label{sec:evol}}
Orbital evolution can drive a satellite into a transient resonance with a normal mode for normal modes with frequency $\omega_\alpha<2\Omega$. However, satellites evolve fast out of the resonance due to the temporally acquired enhanced dissipation. In a planet where normal modes assume a spectrum fixed in time (i.e., no planetary evolution), the observation of an interior--orbital resonance becomes a historical coincidence. 

 Jupiter's dilute core possibly evolves as convective currents erode the dilute core from the top. Due to dilute core erosion, an initially smooth compositional gradient turns sharper over geological timescales, reducing the effective width and extension of the dilute core. For a fixed dilute core outer boundary, our models show that the $_4^2g_1$ mode frequency evolves towards lower frequency as the dilute core becomes narrower (Fig.~\ref{fig:models}). Alternatively to dilute core erosion, double diffusive convection promotes a contrary evolution path for the dilute core, increasing the effective width and extension of the dilute core. The $_4^2g_1$ mode frequency in our models evolves towards higher frequency as the dilute core becomes wider at fixed extension (Fig.~\ref{fig:models}). 

Inertial modes restored by the Coriolis force are unlikely to lead to long--lived interior--orbital resonances because the inertial mode frequency mostly depends on the planet's rotation rate \citep{dewberry2022dynamical}, which rapidly evolves into an essentially constant value a few hundred million years after formation.

\subsection{Principal uncertainties in our proposal}
Our proposed scenario of a long--lived interior--orbital resonance contains large uncertainties. The evolution rate of the dilute core is hard to estimate and it should be high enough as to produce a $_4^2g_1$ mode evolution that exceeds the evolution rate of the Galilean satellites in the absence of resonances (Fig. \ref{fig:evolution}). Not even the sign of the resulting change in the $_4^2g_1$ mode frequency is known, with core erosion and double diffusive convection evolving the dilute core in different directions. Even by formation or evolution, we require an initially extended dilute core with a compositional gradient located at $r\gtrsim0.7R_J$ to obtain an initial $_4^2g_1$ mode frequency that is lower than Io's current tidal frequency. Formation models struggle to produce an extended dilute core that survives convective mixing over the age of the solar system \citep{vazan2018jupiter,liu2019formation,muller2020challenge}. 

The duration of the interior--orbital resonance and the initial dilute core structure are also uncertain. If the dilute core formed with an initial extension $r\sim0.7R_J$, then the interior--orbital resonance needs to be short-lived to produce almost negligible migration for Io over $\sim4.5$ Gyr. If the initial dilute core reached $r\sim0.9R_J$ and shrank or became broader over time, the interior--orbital resonance could be as old as $\sim1.5$ Gyr assuming a resonant migration rate of $\sim 10$ cm/yr (for a compilation of current estimates, see Table 1 in \cite{lainey2009strong}) and negligible migration when out of resonance. In the latter scenario of an initial dilute core with $r\sim0.9R_J$, Io's semimajor axis can expand at most  $\sim 2 R_J$ after formation, which requires Io's formation to occur at $\sim4R_J$. 

Finally, interior--orbital resonances become less likely to occur with high order $_4^2g_n$ modes due to lower coupling with the tidal potential but cannot be ruled out. An interior--orbital resonance with a high order $_4^2g_n$ mode in a less extended dilute core ($r\lesssim 0.7R_J$) could alternatively explain Juno's $k_{42}$ and also be maintained over geological timescales. In this scenario, Io is free to migrate beyond $\sim2R_J$ and the initial dilute core may extend below $r\sim0.9R_J$. 

\subsection{Future tidal and seismological observables of an interior--orbital resonance in Jupiter--Io}
Our proposed scenario of an interior--orbital resonance can be tested against future seismological observations of Jupiter's normal modes and to a lesser extent by the Juno extended mission. We predict that Europa should raise nonresonant tides on Jupiter following the hydrostatic $k_{42}=4.4$ \citep{wahl2020equilibrium}. In a scenario alternative to our proposed interior--orbital resonance, the $\Delta k_{42}=-11\%$ may come from a unknown nonresonant effect and should equally apply to all satellites, leading to the prediction $k_{42}=3.9$ for Europa. At the end of the Juno extended mission, the Europa $k_{42}$ Juno observation will reach an uncertainty $\sigma\approx0.4$ (Luciano Iess, personal communication, 2021 September 26), perhaps allowing us to discriminate between an interior--orbital resonant and a nonresonant model.
Ultimately, we require seismological observations to robustly identify the spectrum of Jupiter's normal modes \citep{gaulme2011detection}. In particular, we predict the $_4^2g_1$ mode frequency to be near Io's current tidal frequency $\omega\approx 270$ $\mu$Hz.

\subsection{He rain}

We discard Jupiter's He rain layer as a potential contributor to the $7\sigma$ discrepancy in $k_{42}$ observed by Juno. The He rain layer in Jupiter produces $g$--modes that oscillate with a frequency that is much lower than the tidal frequency of the Galilean satellites. The immiscibility of He in H produces a He gradient that starts with $Y\approx0.24$ at $r\approx0.85R_J$ and ends with $Y\approx0.28$ at $r\approx0.75R_J$. In this cavity, the $_2^2g_1$ mode oscillates with frequency $\omega_g\approx 0.54\Omega$ and the $_4^2g_1$ mode with frequency  $\omega_g\approx0.87\Omega$, both much lower than the lowest tidal frequency among the Galilean satellites $|\omega|\approx1.5\Omega$ (Io). Higher order He rain  $g$--modes oscillate at even lower frequency (Equation (\ref{eq:gmode})). In the context of gas giant planet evolution, the He rain layer forms near the atmosphere and migrates inward in time as the planet cools down. According to Equation (\ref{eq:gmode}), consequently, the He rain layer hosted $g$--modes with even lower frequencies in the past.

\subsection{Jupiter's dilute core and the dynamo region}
In general, a dynamo region requires to be convectively unstable to produce a magnetic field. A dilute core region promotes vertical stratification of the fluid, presumably shutting down any potential dynamo activity. Recent analysis of Juno observations suggests that Jupiter hosts a dynamo region at $\sim0.8R_J$ capable of reproducing Jupiter’s magnetic field \citep{connerney2021new}. Our extended dilute core model extends up to $\sim0.7R_J$, allowing convection to occur within a layer of metallic hydrogen trapped between the top of the dilute core and the bottom of He rain.

\section{Conclusions}
We used simple Jupiter dilute core models to calculate the fractional dynamical correction to Jupiter's hydrostatic Love number $k_{42}$ from the tidal excitation induced by the Galilean satellites. After considering previously understood  dynamical effects, Juno's $k_{42}$ observation at PJ17 requires an additional fractional correction $\Delta k_{42}\approx -11\%$ to reconcile a $7\sigma$ disagreement with the hydrostatic $k_{42}$. Our results suggest that the required $\Delta k_{42}$ can be produced by an interior--orbital resonance between Io and a $_4^2g_1$ mode (i.e., $\ell,m,n=4,2,1$, a $g$--mode with one radial node in its eigenvector) trapped in Jupiter's dilute core. The tidal dissipation in the $_4^2g_1$ mode is limited to $Q_4\lesssim1000$ or the dynamical tide is damped below the required $\Delta k_{42}\approx -11\%$. Our simple dilute core model achieves a $_4^2g_1$ mode in close resonance with Io's orbital motion when it extends as far as $r\gtrsim 0.7R_J$, a dilute core extension previously suggested in the analysis of zonal gravity data recorded by Juno. Less extended dilute core models could explain Juno's observed $k_{42}$ invoking a resonance of Io with higher order $_4^2g_n$ modes (i.e., additional radial nodes in the mode eigenvector), a possible scenario that cannot be ruled out.

To avoid invoking a historical coincidence in explaining Juno's observation, we propose a scenario where the $_4^2g_1$ mode evolves roughly at a similar rate compared to the rate of Io' current orbital migration, conforming a state of resonant locking that allows the invoked resonance to remain active over geological timescales. We require a tidal dissipation $Q_4\lesssim1000$ to maintain the aforementioned state of resonant locking. Our proposed self--consistent scenario depends on largely unconstrained assumptions about the long--term evolution of Jupiter's dilute core. On the short term, Juno may provide the first test for our proposal from the end--mission observation of Jupiter's $k_{42}$ associated to tides raised by Europa. On a longer term, future seismological observations of Jupiter's normal modes can test the validity of our proposal from an observation of our predicted $_4^2g_1$ mode frequency ($\omega_g\approx 270$ $\mu$Hz).


\begin{acknowledgments}
We acknowledge the support of NASA's Juno mission. We are grateful for the constructive comments of Tim Van Hoolst and one anonymous referee. We benefited from constructive discussions with James Fuller, Christopher Mankovich, Luciano Iess, and Daniele Durante.
\end{acknowledgments}

%

\vspace{5mm}


\software{Matplotlib \citep{hunter2007matplotlib}, GYRE \citep{townsend2013gyre}
          }



\appendix

\section{Dynamical Love numbers of gas giant planets\label{app:dynamic}}

The dynamical Love number describes dynamical tides where inertial effects in the response of the planet are taken into account. The associated tradional equation of motion corresponds to $F = M\dot{v}$, where $F$ includes the tidal forcing and self--gravitation of the tidally displaced mass, $M$ the tidally displaced mass, and $v$ the tidal flow. Alternatively, the hydrostatic Love number describes an instantaneous response where the equation of motion to solve reduces to the traditional $F = 0$. In a gas giant planet, the linearly perturbed equation of motion for dynamical tides becomes \citep{idini2021dynamical}
\begin{equation}
      -i\omega \mathbf{v} + 2\mathbf{\Omega} \times \mathbf{v} = - \frac{\nabla p'}{\rho} + \frac{\rho'}{\rho^2}\nabla p + \nabla\tilde{\phi}' \textnormal{,}
      \label{eq1:momentum}
\end{equation}
where $\omega$ is the tidal frequency, $\bm{\Omega}$ is the spin rate of the planet, $p$ is pressure, $\rho$ is density, and $\tilde{\phi}$ is the resulting tidal potential after adding the tidal forcing potential and the self-gravitation potential. The primes represent Eulerian linear perturbations. 

Alternatively to solve the tidal equation of motion (Equation~\ref{eq1:momentum} coupled to Poisson's and continuity equations; for an example, see \cite{idini2021dynamical}), the dynamical Love number can be obtained from considering the dynamical response of periodically excited harmonic oscillators with oscillation frequency $\omega_0$ \citep{idini2021dynamical}
\begin{equation}
    \Delta k = \frac{\omega^2}{\omega^2_0-\omega^2}\textnormal{,}
    \label{eq2:dk2}
\end{equation}
where $\Delta k$ represents the fractional correction to the hydrostatic Love number due to dynamical effects. In Equation (\ref{eq2:dk2}), the frequency $\omega_0$ represents the $f$--mode oscillation frequency of the planet, which is forced by a periodic loading with frequency $\omega$  associated to the gravitational pull of the companion satellite. 

Several additional effects complicate a practical use of Equation (\ref{eq2:dk2}), which is qualitatively illuminating but fails at delivering useful predictions. In Jupiter and Saturn, fast rotation introduces the Coriolis effect as an important new term in the equation of motion (i.e., the term $2\bm{\Omega} \times \bm{v}$ in Equation~\ref{eq1:momentum}). Even in this case, the dynamical Love number can still be represented by periodically forced normal modes, but the new dynamics change the tidal motion in important ways. The $f$--mode dynamical response changes in sign and amplitude \citep{idini2021dynamical}, and a new set of modes restored by the Coriolis force join the already existent $p$-- and $f$--modes \citep{dewberry2022dynamical}. 

When rotation is treated as a linear perturbation, we can write the dynamical Love number as \citep{Lai_2021}
 \begin{equation}
     k_{\ell m}\simeq \left(\frac{4\pi}{2\ell + 1}\right)\sum_\alpha \frac{\mathcal{Q}_\alpha ^2}{\omega_\alpha^2-(mC_\alpha \Omega+\omega)^2}\textnormal{,}
     \label{eq:dlai2}
 \end{equation}
 where the sum over $\alpha$ represents a sum over all normal modes trapped inside the planet, $\omega_\alpha$ is the mode frequency, $\mathcal{Q}_\alpha$ is the coupling integral defined in Equation (\ref{eq:integral_coupling}), $C_\alpha$ is the rotational coefficient defined in Equation (\ref{eq:rotational_coeff}), $\ell$ is degree, and $m$ is azimuthal order. 
 
 The hydrostatic Love number emerges from $f$--modes in Equation (\ref{eq:dlai2}) when $\omega=0$. Other modes (i.e., $p$--, $g$--, and inertial modes restored by Coriolis) weakly couple to the tidal forcing (i.e., they have a small $\mathcal{Q}_\alpha$), thus only contribute to the dynamical Love number when the mode frequency $\omega_\alpha$ approaches the forcing frequency $\omega$. The dynamical contribution from inertial modes restored by Coriolis requires a nonperturbative treatment of rotation, which complicates the numerical calculation of mode properties but preserves the main idea behind Equation (\ref{eq:dlai2}) \citep{dewberry2022dynamical}.
 

\section{Resonant locking applied to a planet--satellite system \label{app:resonant}}

In a state of resonant locking, the migration rate $\dot{a}$ of the satellite depends only on the planetary evolution of the central planet \citep{fuller2016resonance},
\begin{equation}
    \frac{\dot{a}}{a} = \frac{2}{3}\left( \frac{\Omega}{\omega_s}\left(\frac{\dot{\omega}_\alpha}{\omega_\alpha}-\frac{\dot{\Omega}}{\Omega}\right) - \frac{\dot{\omega}_\alpha}{\omega_\alpha} \right)\textnormal{,}
\end{equation}
where $\Omega$ is the spin--rate of the planet, $\omega_s$ is the orbital frequency of the satellite, and $\omega_\alpha$ is the oscillation frequency of the resonant mode trapped inside the planet. Resonant locking extends the lifetime of interior--orbital resonances to geological timescales, diminishing the secular tidal $Q$ of the tidal bulge associated to the resonant mode and modifying the secular amplitude of dynamical tides.

The basic conditions required to maintain a state of resonant locking are the following \citep{fuller2016resonance}:
\begin{enumerate}
    \item The resonance must greatly increase the amount of tidal dissipation after it kicks in, so that satellite migration accelerates when $\omega\approx \omega_\alpha$.
    \item The resonance must migrate in the same direction of satellite migration. For a satellite migrating outwards, the mode frequency should increase in time to encounter the also increasing tidal frequency $\omega=m(\Omega-\omega_s)$, where $m$ is the azimuthal order and $\omega_s$ diminishes as the orbit expands. 
    \item Tidal dissipation at resonance must accelerate the satellite migration to match the evolution of the resonant mode. If satellite migration cannot match the mode evolution $\dot{\omega}_\alpha$ before saturation of the resonance, the resonance migrates past the orbital frequency and resonant locking fails. 
\end{enumerate}

\section{A simplified equation of state of mixtures\label{app:mixture}}

We compute the pressure $p$ in a mixture of H-He fluid with density $\rho$ and heavy elements with density $\rho_z$ using volume additivity of the individual constituents (Equation (\ref{eq:specificvol})). For H-He fluid following an $n=1$ polytropic equation of state with a He cosmic abundance, the pressure follows
\begin{equation}
    p = K \rho^2f^2\textnormal{,}
    \label{eq:pmix}
\end{equation}
where the radial function $f(r)$ is
\begin{equation}
    f(r) = \frac{1-Z}{1-\frac{\rho}{\rho_z}Z}
\end{equation}
The ratio $\rho/\rho_z$ becomes small near the atmosphere but non-negligible near the core region. To a good approximation, the pressure-density relation for an adiabatic distribution of 'rocky' heavy elements follows \citep{hubbard1989optimized}:
\begin{equation}
    p \approx \left(\frac{1.4}{1000}\right)\rho_z^{4.4}\textnormal{,}
    \label{eq:rocky}
\end{equation}
with $p$ in Mbar and $\rho_z$ in g/cm$^3$. Near the center of Jupiter, pressure reaches $p\sim \mathcal{G}M_J^2/4R_J^4\sim 30 $Mbar and fluid density $\rho\sim 4.4$ g/cm$^3$, which leads to a ratio $\rho/\rho_z\sim 0.4$. We approximate the ratio $\rho/\rho_z$ by treating $\rho$ as the density of pure H-He fluid and $\rho_Z$ as the density of 'rocky' heavy elements (Equation \ref{eq:rocky}) subjected to the hydrostatic pressure obtained from a body made of pure H-He fluid (i.e., $p = K\rho_c^2 j_0^2(kr)$)
\begin{equation}
    \frac{\rho}{\rho_z} \approx 0.42 \left(j_0(kr)\right)^{0.6} \textnormal{.}
\end{equation}
We can now rewrite $f(r)$ as
\begin{equation}
    f(r) = \frac{1-Z}{1- 0.42  \left(j_0(kr)\right)^{0.6} Z } \textnormal{.}
\end{equation}

The hydrostatic density profile of the mixture comes from solving a differential equation similar to Lane-Emden's equation. After defining the auxiliary variable $\tilde{\rho} = \rho f$, hydrostatic equilibrium imposes
\begin{equation}
    \frac{1}{\rho} \frac{\partial p}{\partial r} = 2Kf\frac{\partial \tilde{\rho}}{\partial r}=-g \textnormal{.}
\end{equation}
After substituting the previous equation into the resulting equation from differentiating with respect to $r$, we reach a second-order differential equation on $\tilde{\rho}$,
\begin{equation}
    \frac{\partial^2 \tilde{\rho}}{\partial r^2} + \left(\frac{\partial\log f}{\partial r} +\frac{2}{r}\right)\frac{\partial \tilde{\rho}}{\partial r} + \left(\frac{k}{f}\right)^2\tilde{\rho} = 0\textnormal{.}
    \label{eq:edo2rho}
\end{equation}
The usual boundary conditions $\rho= \rho_c$ and $\rho'=0$ at the center lead to a new set of boundary conditions
\begin{equation}
    \tilde{\rho}(0) = f(0)\rho_c=\left(\frac{1-Z_c}{1-0.42Z_c}\right)\rho_c\textnormal{,}
\end{equation}
\begin{equation}
    \left.\frac{\partial \tilde{\rho}}{\partial r}\right|_{r=0} = 0\textnormal{.}
\end{equation}
We solve Equation (\ref{eq:edo2rho}) numerically after normalizing $\tilde{\rho}$ by the central density $\rho_c$. We use the result to calculate $\rho_c$ from Jupiter's mean density $\bar{\rho}_J$, according to
\begin{equation}
    \bar{\rho}_J = \left(\frac{3}{R_J^3}\right)\int^{R_J}_0 \rho r^2 dr \textnormal{.}
\end{equation}


\bibliography{sample631}{}
\bibliographystyle{aasjournal}



\end{document}